\documentstyle[preprint,aps]{revtex}
\begin{document}

\draft

\title{Valence Bond Mapping of Antiferromagnetic Spin Chains}

\author{J.~Dukelsky$^{1}$ and S.~Pittel$^{2}$}
\address{
$^{1}$ Instituto de Estructura de la Materia, Consejo Superior de\\
Investigaciones Cientificas, Serrano 123, 28006 Madrid, Spain \\
$^{2}$Bartol Research Institute, University of Delaware,\\
Newark, DE 19716 USA }

\date{\today}

\maketitle

\begin{abstract}
Boson mapping techniques are developed to describe valence bond correlations
in quantum spin chains. Applying the method to the alternating bond
hamiltonian for a generic spin chain, we derive an analytic expression for
the transition points which gives perfect agreement with existing Density
Matrix Renormalization Group (DMRG) and Quantum Monte Carlo (QMC)
calculations.
\end{abstract}

%***************** ABSTRACT ************************

\begin{center}
\end{center}

\newpage
Antiferromagnetic spin chains have been the subject of intense interest in
recent years, largely due to the conjecture by Haldane \cite{R2} that chains
built from integer spins should exhibit a gap in their energy spectrum. The
existence of this gap, originally predicted on the basis of simple
field-theoretic considerations, was subsequently confirmed experimentally
\cite{R3}.

Much insight into the properties of antiferromagnetic spin chains has been
provided by simple models. The field-theoretic non-linear sigma model (NLSM)
\cite{R2,R4}, for example, provided the original motivation for Haldane's
conjecture. Equally illuminating insight into the diverse properties of spin
chain systems has been provided by the introduction of the Valence Bonds Solid
(VBS) state \cite{R5}.

Detailed descriptions of these systems, however, have depended on very
complex numerical analyses, using either Quantum Monte Carlo (QMC) simulations
\cite{R6} or Density-Matrix Renormalization Group (DMRG) methods \cite{R7}. 
These ``exact'' treatments provide striking confirmation of the features 
conjectured by Haldane.

Ideally, it would be nice to have a simpler method for a reliable quantitative
treatment of quantum spin chains.  Valence bonds provide a physically--motivated
starting point for such a description. The VBS is the exact ground state of
specific spin-chain hamiltonians involving quadratic and quartic terms,
suggesting that wave functions constructed in terms of valence bonds might be
good trial states more generally. Unfortunately, such wave functions are still
too complex, for reasons to be discussed later, to be useful in a variational
analysis. In the present work, we propose a method that permits valence 
bonds to be used efficiently in variational calculations. We also describe 
the application
of this method to study the phase transitions in spin chains governed by the
alternating bond hamiltonian \cite{R8}.

A useful starting point for the introduction of valence bonds is through the
Schwinger boson realization of the spin algebra \cite{R9}. The basic idea is
to introduce a set of boson creation and annihilation operators $\gamma
_{i,\sigma }^{\dagger }$ and $\gamma _{i,\sigma }$, respectively. These
operators create and annihilate, respectively, a spin-$\frac {1}{2}$ boson with
spin projection $\sigma =+\frac {1}{2}$ (denoted +) or $\sigma =-\frac {1}{2}$
(denoted --) at site $i$. A spin-S system would then involve $2S$ Schwinger
bosons on each site.

Typical spin-chain hamiltonians involve spin-spin interactions between
nearest neighbors. It is natural therefore to consider states built up in
terms of bonds reflecting these correlations. This is the basic idea behind the
introduction of valence bonds.  When dealing with an antiferromagnetic spin
chain, the key correlations involve nearest neighbors in spin-singlet states,
which can be represented by the singlet bond
\begin{equation}
\Gamma _i^{\dagger }=\frac 1{\sqrt{2}}(\gamma
_{i,+}^{\dagger }\gamma _{i+1,-}^{\dagger }-\gamma _{i,-}^{\dagger }\gamma
_{i+1,+}^{\dagger }) ~.
\end{equation}
In terms of these singlet bonds, the VBS
ground state for a spin-S chain (S an integer) is given by

\begin{equation}
|VBS>~=~\prod_{i=1,N}(\Gamma _i^{\dagger })^S|0>  ~.
\end{equation}

The state $|VBS>$ is the exact ground eigenstate of the
hamiltonian
\begin{equation}
H=\sum_i\left[ S_i\cdot S_{i+1}+\frac 13\left( S_i\cdot S_{i+1}\right)
^2\right]
\end{equation}
involving quadratic and quartic spin operators. For a general hamiltonian,
however, it is not an eigenstate. Furthermore, when not an exact eigenstate, it
is not an especially useful trial state. The reason is that the singlet bonds
$\Gamma _i^{\dagger }$ from which it is built are not bosons, i.e., the
corresponding operators do not satisfy boson commutation relations.

This is a familiar problem in many branches of physics. A standard approach
to problems of this type, involving dominant pair correlations, is to
implement a {\it boson mapping} \cite{R10}.

The general idea of a boson mapping is to replace the original problem
involving pair degrees of freedom (in this case, singlet bonds) and the
true hamiltonian of the system by an {\it equivalent} problem involving real
bosons and an appropriate effective hamiltonian for these bosons. All of the
exchange effects between the constituents is transferred to the effective
hamiltonian, in a mathematically rigorous way guaranteed to preserve the
physics of the original problem.

In more detail, we wish to {\it replace} the valence bonds $\gamma
_{i,\sigma _1}^{\dagger }\gamma _{i+1,\sigma _2}^{\dagger }$ by bosons 
$B_{i,\sigma _1\sigma _2}^{\dagger }$, which fulfill exact bosonlike
commutation relations:

\begin{equation}
\left[ B_{i,\sigma _1\sigma _2~},~B_{j,\sigma _3\sigma _4}^{\dagger }\right]
=\delta _{i,j~}\delta _{\sigma _1,\sigma _3}~\delta _{\sigma _2,\sigma _4} ~.
\end{equation}

There are many ways to implement such a replacement. In the Generalized
Holstein-Primakoff (GHP) approach \cite{R10}, which we follow here, the
mapping is defined by imposing the requirement that all quadratic operators
in the original space are mapped in such a way as to preserve their
commutation relations. More specifically:

\begin{itemize}
\item  The boson image of quadratic operators are assumed to be given by
{\it Taylor-series expansions:}

\begin{equation}
F_B=F^{(0)}+F^{(1)}+F^{(2)}+.....
\end{equation}

\item  The terms in these Taylor expansions are obtained via the condition
that any commutation rule $\left[ A,B\right] =C$ between the original set of
quadratic operators must be preserved at each order of the expansion:
\begin{eqnarray}
\left[ A^{(0)},B^{(0)}\right] &~~=~~& C^{(0)}  \nonumber \\
\left[ A^{(0)},B^{(1)}\right] +\left[ A^{(1)},B^{(0)}\right] &~~=~~& C^{(1)}
\nonumber \\
&.....&
\end{eqnarray}
\end{itemize}

We should note here that boson mappings have typically been applied to
systems of interacting fermions. However, there is no fundamental difficulty
in applying them to systems of interacting bosons \cite{R11}, as in the
problems under discussion.

We have succeeded in building an appropriate boson mapping of the
Holstein-Primakoff type for valence bonds. In the present discussion, we
simply present the relevant mapping equations, leaving more detailed
discussion of the formalism to a subsequent publication.

The full algebra of quadratic operators in the Schwinger boson space
includes both particle-hole (p-h) operators of the form 
$\gamma^{\dagger}_{i\sigma}\gamma_{j\sigma^{\prime}}$ 
and particle-particle operators of the
form $\gamma^{\dagger}_{i\sigma}\gamma^{\dagger}_{j\sigma^{\prime}}$ and 
$\gamma_{i\sigma}\gamma_{j\sigma^{\prime}}$.

For the purpose of treating the dynamics of spin-chain hamiltonians, we only
need to know how to map the on-site p-h operator 
$\gamma^{\dagger}_{i,\sigma}\gamma_{i,\sigma^{\prime}}$ 
and the p-p operators involving bonds
between neighboring sites, $\gamma^{\dagger}_{i,\sigma}
\gamma^{\dagger}_{i+1,\sigma^{\prime}}$ and $\gamma_{i,\sigma}
\gamma_{i+1,\sigma^{\prime}}$. The relevant images through first-order 
in the Taylor series
expansion are as follows:

The boson image of the on-site particle-hole
operator is:
\begin{equation}
(\gamma _{i,\sigma _1}^{\dagger }\gamma _{i,\sigma
_2})_{B}=\sum_{\sigma_3}\left\{ B_{i,\sigma _1\sigma _3}^{\dagger }B_{i,\sigma
_2\sigma _3}+B_{i-1,\sigma _3\sigma _1}^{\dagger }B_{i-1,\sigma _3\sigma
_2}\right\} ~.
\end{equation}
Note that the on-site p-h operator maps exactly onto a p-h operator in the
ideal boson space, with no need for a series expansion.

The p-p operators involving nearest neighbor bonds do require infinite
series expansions. The lowest (zeroth) order images are straightforwardly
given by

\begin{equation}
\left( \gamma _{i,\sigma _1}^{\dagger }\gamma _{i+1,\sigma _2}^{\dagger
}\right) _{B}^{(0)}=B_{i,\sigma _1\sigma _2}^{\dagger }
\end{equation}
and
\begin{equation}
\left( \gamma _{i,\sigma _1}\gamma _{i+1,\sigma _2}\right)
_{B}^{(0)}=B_{i,\sigma _1\sigma _2}~.
\end{equation}

The first-order images are more complex, as they provide the first
reflection of exchange effects of the (Schwinger boson) constituents between
neighbor bonds. The required first--order image that fulfills the second line of
(6) is given by

\begin{eqnarray}
\left( \gamma _{i,\sigma _1}^{\dagger }\gamma _{i+1,\sigma _2}^{\dagger
}\right) _{B}^{(1)}=\frac {1}{2}\sum_{\sigma_3,\sigma_4} && \left\{  B_{i,\sigma
_1\sigma _3}^{\dagger }B_{i,\sigma _4\sigma _1}^{\dagger }B_{i,\sigma _4\sigma
_3}+B_{i,\sigma _1\sigma _3}^{\dagger }B_{i+1,\sigma _2\sigma _4}^{\dagger
}B_{i+1,\sigma _3\sigma _4} \right. \nonumber \\
&&~~~~\left. +  B_{i,\sigma _3\sigma _2}^{\dagger }B_{i-1,\sigma
_4\sigma _1}^{\dagger }B_{i-1,\sigma _4\sigma _3}\right\} ~.
\end{eqnarray}
The corresponding image associated with annihilation of a bond between
nearest neighbors is obtained from (10) by hermitian conjugation.

In fact, closing the algebra to first order requires inclusion of the p-h
operator between next-to-nearest neighbor ($i$ with $i+2$) sites. We will
not discuss this any further here, as it does not impact on the analysis or
results to follow.

There is one further complication that should be noted before considering the
application of these methods. The mapping equations 
given above can be applied in
several different ways to a given spin-chain hamiltonian. 
One possibility is to
express the two-body interaction entering the hamiltonian in p-p form
($\gamma^{\dagger} \gamma^{\dagger} \gamma \gamma$) and to map 
using eqs. (8-10).
Alternatively, the hamiltonian could first be transformed into p-h form
($\gamma^{\dagger} \gamma \gamma^{\dagger} \gamma$) and then mapped 
with eq. (7).
A third possibility, of course, is to map part of the hamiltonian in 
p-h form and
part in p-p form. Were we to do the resulting analysis {\it exactly} in the 
ideal boson
space, all such approaches would be equivalent. In
variational treatments, on the other hand, it is essential to map the
hamiltonian in such a way as to maintain the key correlation effects.

As a specific application of these methods, we consider a spin-chain system
governed by the alternating bond hamiltonian \cite{R8},

\begin{equation}
H(\alpha )~=~\sum_{i=1}^N[1-(-)^i\alpha ]S_i\cdot S_{i+1} ~.
\end{equation}
This system has been studied extensively, especially with regards to its
dimer phase transitions.

The variational treatment we will apply to this system is based on singlet bonds
only. Thus, following our earlier remarks, we must map the hamiltonian so as
to most efficiently reflect these bonds. The appropriate separation is:

\begin{equation}
H(\alpha )~=H_1(\alpha )~+~H_2(\alpha )~,
\end{equation}
where
\begin{eqnarray}
H_1(\alpha )~ &=&~\sum_{i=1}^N[1-(-)^i\alpha ]S_i^zS_{i+1}^z  \nonumber \\
&=&\frac 14\sum_{i=1}^N[1-(-)^i\alpha ]\sum_\sigma \sigma \ \gamma
_{i,\sigma }^{\dagger }\gamma _{i,\sigma }\sum_{\sigma ^{\prime }}\sigma
^{\prime }\gamma _{i+1,\sigma ^{\prime }}^{\dagger }\gamma _{i+1,\sigma
^{\prime }}    ~,
\end{eqnarray}
and
\begin{eqnarray}
H_2(\alpha )~ &=&~\frac 12\sum_{i=1}^N[1-(-)^i\alpha
](S_i^{+}S_{i+1}^{-}~+~S_i^{-}S_{i+1}^{+})  \nonumber \\
&=&\frac 12\sum_{i=1}^N[1-(-)^i\alpha ]\sum_\sigma \gamma _{i,\sigma
}^{\dagger }\gamma _{i+1,-\sigma }^{\dagger }\gamma _{i,-\sigma }\gamma
_{i+1,\sigma }  ~.
\end{eqnarray}
We map the term $H_1(\alpha )$, corresponding to $S_i^zS_{i+1}^z$, in p-h
form and the term $H_2(\alpha )$, corresponding to $%
S_i^{+}S_{i+1}^{-}+S_i^{-}S_{i+1}^{+}$, in p-p form.

Applying the mapping in this way to the alternating bond hamiltonian (11) and
then projecting onto singlet bosons, defined by

\begin{equation}
\sigma_i^{\dagger} ~=~ \frac{1}{\sqrt{2}} [ B^{\dagger}_{i,+-}
-B^{\dagger}_{i,-+}] ~,
\end{equation}
we obtain

\begin{equation}
H_{B}(\alpha)~=~-\frac 14\sum_{i=1}^N[1-(-)^i\alpha ]\left\{ 3\sigma
_i^{\dagger }\sigma _i
+\sigma _i^{\dagger }[\sigma _i^{\dagger }\sigma _i+\sigma
_{i+1}^{\dagger }\sigma _{i+1}+\sigma _{i-1}^{\dagger }\sigma _{i-1}]\sigma
_i\right\} ~.
\end{equation}

The trial wave function we use in our variational description of the
alternating bond spin-chain system is

\begin{equation}
{\displaystyle |\Phi _{n_o,n_e}>~=~\prod_{i(odd)=1}^N\frac{\sigma
_i^{\dagger ~n_o}\sigma _{i+1}^{\dagger ~n_e}}{\sqrt{n_o!n_e!}}|0>}~,
\end{equation}
subject to the Schwinger constraint
\begin{equation}
n_o+n_e~=~2S~.
\end{equation}

This trial wave function reflects the various phases of the system. The
Heisenberg phase corresponds to $n_o=n_e$, with all sites involved in an equal
number of bonds with its nearest neighbors on each side. The corresponding state
is translationally invariant and is the analogue of the VBS state in the ideal
boson space. Increasing $n_o$ or equivalently $n_e$ corresponds to successive
partial dimerization. The cases in which either $n_o=0$ or $n_e=0$ involve
complete dimerization.

Here we focus our analysis on the location of the critical points associated
with a transition from one phase to another. Defining
\begin{equation}
{\cal E }_{n_o,n_e}(\alpha) ~=~ <\Phi_{n_o,n_e} | H_B (\alpha) |
\Phi_{n_o,n_e}> ~,
\end{equation}
the critical points are given by the condition
\begin{equation}
{\cal E }_{n_o,n_e} ~=~ {\cal E }_{n_o \pm 1,n_e \mp 1} ~.
\end{equation}

Straightforward analysis yields for the energy functional,
\begin{eqnarray}
{\cal E}_{n_o,n_e}(\alpha )~=~-\frac N4 &&\left[ (1+\alpha )\{\frac{3n_o}{2}+
\frac{n_o(n_o-1)}2+n_on_e\}\right.  \nonumber \\
&&+\left. (1-\alpha )\{\frac{3n_e}2+\frac{n_e(n_e-1)}2+n_on_e\}\right] ~,
\end{eqnarray}
and for the critical values of $\alpha $,
\begin{equation}
\alpha ~=~\frac{2n_o+1-2S}{2(S+1)}~,~~~~~~~~~n_o=0,1,...,2S-1~.
\end{equation}

In Table 1, we present the results of this analysis for several values of the
spin $S$. We compare the results obtained from our simple analytic formula (22)
for the crossing points with those from ``exact'' 
calculations \cite{R12,R13,R14}
and from the NLSM \cite{R2,R4}. Our simple formula, obtained by using a
first-order boson mapping treatment and a simple product trial state, yields
perfect agreement with the exact results where available. This is to be
contrasted with the NLSM results, which are in much worse agreement. 
Note that we
have also included in the table {\it predictions} for the location of the phase
transition points at higher spins.

The fact that our first-order boson mapping works so well in describing the
location of dimer phase transitions does not mean that higher-order
contributions are of no dynamical importance. Indeed, at the same level of
approximation, the energy per site of the pure $S=1$ Heisenberg lattice 
($\alpha =0$) is calculated to be $-5/4$, whereas the exact result is 
$-1.401...$ \cite{R7}. Thus, there is clearly room for improvement, which
we expect would be provided in part by the next-order contribution of the
boson mapping and in part by fluctuation effects. What is important to
realize, however, is that the method we have outlined provides a systematic
way to incorporate such improvements.

There are several areas in which we expect these methods to be useful in the
future.  Our immediate plan is to generalize the hamiltonian (11) to include
crystal fields and a uniform magnetic field and to study the phase diagram,
excitations, magnetization curves, etc.

This work was supported in part by the DIGICYT (Spain) under contract no.
PB95/0123 and by the National Science Foundation under grant \# PHY-9600445.
Fruitful discussions with Siu Tat Chui and German Sierra are gratefully
acknowledged.

%%%%%%%%%%%%%%%%%%%%%%%%%%%%%%%%%%%%%%%%%%
% Table 1 %%%%%%% Crossing Points %
%%%%%%%%%%%%%%%%%%%%%%%%%%%%%%%%%%%%%%%%%%
\begin{table}[tbp]
\caption{Location of the crossing points for the alternating bond spin
chain. In addition to the results of the present analysis, we present
results from the non-linear sigma model (NLSM) and from ``exact" numerical
solution. In the case of the exact analyses, only the nonnegative crossing
points are shown.}
\begin{tabular}{lcccc}
& $S$ & Exact & Present Results & NLSM \\
\tableline
& 1 & 0.25 $\pm$0.01 & $\pm$1/4 & $\pm$1/2 \\
& 3/2 & 0.~, ~0.42 $\pm$0.02 & 0~, ~$\pm$2/5 & 0~, ~ $\pm$2/3 \\
&2&0.05$<\alpha<$0.3~,~ 0.5$<\alpha<$0.6& $\pm$1/6~,~
$\pm$1/2& $\pm$1/4~,~$\pm$3/4 \\
& 5/2 & --- & 0~, ~$\pm$1/3~,~$\pm$2/3 & 0~, ~
$\pm$2/5~,~ $\pm$4/5 \\ & 3 & --- & $\pm$1/7~,~$\pm$3/7~,~$\pm$5/7 &
$\pm$1/6~,~$\pm$1/2~,~$\pm$5/6 \\
\end{tabular}
\tablenotetext[1]{The exact results for $S=1$ are from Ref. 11, 
those for $S=3/2$
are from Ref. 12, and those for $S=2$ are from Ref. 13}
\end{table}

\end{document}